# Really Need Recursion? Revisiting One-pass Registration with Multi-level Multi-space Contextual Guidance

Yue Lin[1,2][0009-0008-7758-4530], Shenghai Wu[2][0009-0001-9473-136X], and Mingyuan Meng[1](✉)[0000-0002-9562-1613]

[1] Zhongguancun Academy, Beijing, China.
[2] Research Institute, Newland Digital Technology, Fuzhou, China.
mengmingyuan@bza.edu.cn

**Abstract.** Deformable image registration is a fundamental challenge in medical image analysis. Recent deep learning-based methods typically adopt either one-pass architectures that complete registration in a single network iteration, or recursive architectures that refine registration through multiple passes. While recursive methods often improve accuracy, they incur substantial computational overhead and slower inference speed. In this paper, we challenge the necessity of recursion for high-performance deformable registration and propose **M2CG**, a one-pass registration network with **M**ulti-level **M**ulti-space **C**ontextual **G**uidance. It performs multi-step progressive pyramid registration within a single iteration, guided at each step by contextual cues from three levels (previous, current, and subsequent steps) and four spaces (image space, image feature space, segmentation space, and segmentation feature space). Our M2CG supports both unsupervised (without any labels) and semi-supervised (with segmentation labels for partial training data) registration, and has achieved state-of-the-art performance across two well-benchmarked inter-/intra-patient medical registration tasks with eight public datasets. Our findings suggest that the current reliance on recursion may stem from the absence of sufficient contextual cues that comprehensively characterize the remaining distance to the target alignment.

## 1 Introduction

Medical image registration aims to spatially align medical images acquired from different patients, time points, or scanners, which has been an active research focus for decades [1, 2]. Deformable medical image registration finds a dense (pixel-wise) nonlinear spatial transformation between a pair of images, allowing the two images to be spatially aligned after warping. Traditional methods typically formulate deformable registration as a time-consuming iterative optimization problem [3, 4], while recent deep registration methods based on deep neural networks have reformulated it as a fast end-to-end mapping from image pairs to spatial transformations, achieving superior registration performance with remarkably faster speed [2].

Existing deep registration methods are typically divided into two categories: (i) one-pass registration, which completes alignment in a single network iteration [5-10], and (ii) recursive registration, which refines alignment through multiple passes [11-16]. The relative merits of these two categories have been a subject of long-standing debate. Early methods, including the widely used VoxelMorph [5] and TransMorph [6], used Unet-style architectures for fast one-pass registration, with an initial focus on achieving higher speed than iterative optimization-based methods. In the same period, RCN [11] pioneered cascading multiple networks (e.g., VoxelMorph) to perform recursive registration for improved accuracy. Subsequent studies cascaded Laplacian pyramid networks [12] or iteratively ran a single network [13] for recursive registration, thereby alleviating the computational cost of RCN while maintaining superior alignment accuracy over contemporary one-pass methods. However, the desire for fast, efficient one-pass registration is inherent, spurring substantial studies focusing on designing high-performance one-pass registration methods [7-10, 35]. For example, NICE-Net [7] introduced a non-iterative coarse-to-fine network that completes multi-step pyramid registration within a single network iteration. It achieved superior alignment accuracy over earlier recursive methods with only one-pass computational costs, and similar concepts have since been widely adopted in current one-pass methods relying on progressive pyramid registration architectures [8-10]. Nevertheless, even within advanced pyramid registration architectures, applying recursion at each pyramid step remains beneficial for further improving alignment accuracy [14-16].

Throughout these developments, there seems to be a prevailing belief that recursion can serve as a panacea to improve registration accuracy, albeit with substantial computational overhead. For instance, a recent study [17] leveraged a recursive registration network as a high-performing teacher to distill knowledge into a lightweight one-pass student registration network, reflecting this view of recursion versus one-pass trade-off. So far, few studies have critically examined why and when recursion is really needed for deformable registration. For current one-pass pyramid registration architectures, step-wise contextual information is crucial for characterizing the remaining distance to the final target alignment and guiding the optimization direction of each registration step [18]. We hypothesize that, given sufficient contextual cues to comprehensively characterize the status of each registration step, pyramid architectures can capture deformations via one-pass progressive registration alone; Otherwise, recursion is needed to compensate for vague guidance cues by iterative searching. Thus, we suggest that the current reliance on recursion may stem from the absence of such informative contextual guidance. Moreover, it is worth noting that, while recursion aids in approximating to the final target, it may also introduce accumulated errors.

To validate the above hypothesis, we propose **M2CG**, a one-pass pyramid registration network with **M**ulti-level **M**ulti-space **C**ontextual **G**uidance. It performs progressive pyramid registration within a single network iteration, with each registration step guided by contextual cues from three levels (previous, current, and subsequent) and four spaces (image space, image feature space, segmentation space, and segmentation feature space). To the best of our knowledge, this is the first study to empirically investigate the necessity of recursion under sufficient contextual guidance in deformable registration, offering the potential to influence the designs of future deep registration methods. M2CG supports both unsupervised (without any labels) and semi-supervised (with segmentation labels for partial training data) registration, and we evaluated it on two well-benchmarked inter-/intra-patient medical registration tasks with eight public datasets under both settings.

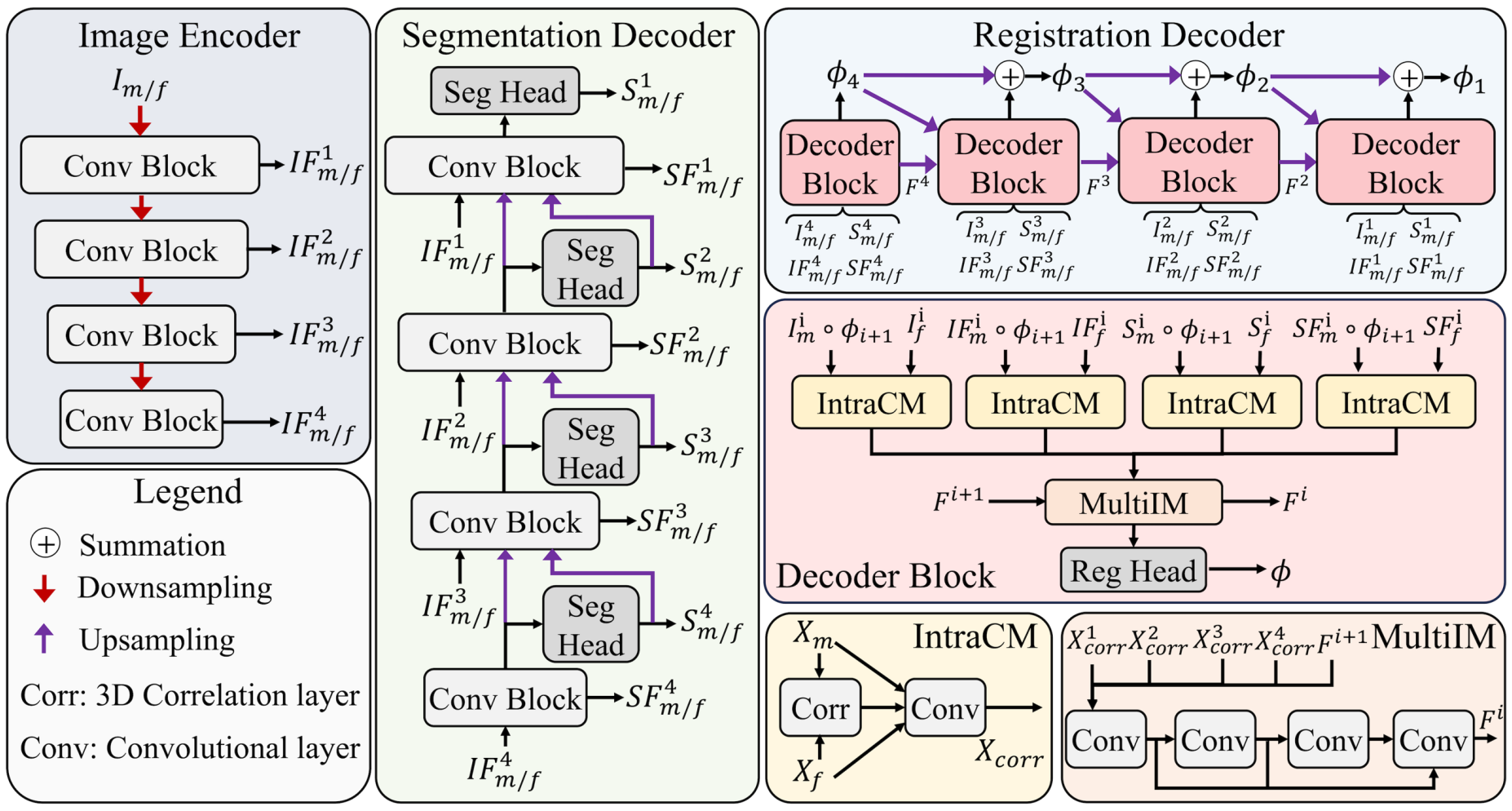

**Fig. 1.** Illustration of the proposed M2CG, consisting of a image encoder, a segmentation decoder, and a registration decoder with multi-level multi-space contextual guidance.

## 2 Method

Deformable image registration finds a spatial transformation $\phi$ between a moving image $I_m$ and a fixed image $I_f$, so that the warped image $I_{m\circ\phi} = I_m \circ \phi$ is spatially aligned with $I_f$. In this study, we assume $I_m$ and $I_f$ are two single-channel, grayscale 3D volumes. The $\phi$ is parameterized as a displacement field, and we parametrized the image registration problem as a function $\mathcal{R}_\theta(I_f, I_m) = \phi$, with $\theta$ as learnable parameters. The function $\mathcal{R}_\theta$ is implemented as the proposed M2CG in this study.

Fig. 1 illustrates the overall architecture of M2CG, consisting of a image encoder for image feature extraction (Section 2.1), a segmentation decoder that performs auxiliary segmentation for semi-supervised registration (Section 2.1), and a registration decoder that performs one-pass pyramid registration with multi-level multi-space contextual guidance (Section 2.2). The learnable parameters $\theta$ can be optimized through unsupervised or semi-supervised learning (Section 2.3).

### 2.1 Image Encoder and Segmentation Decoder

A image encoder, shared between the fixed and moving images, is used to extract multi-scale image features. The encoder comprises four sequential convolutional blocks, and each block consists of two 3×3×3 convolutions. Between blocks, average pooling with stride 2 halves the spatial resolution while doubling the feature channels. This yields a four-level image feature pyramid, denoted as $\{IF^1_{m/f}, IF^2_{m/f}, IF^3_{m/f}, IF^4_{m/f}\}$ for $I_{m/f}$.

In the semi-supervised setting, a segmentation decoder progressively predicts segmentation masks from multi-scale image features. This decoder first takes $F^4_{m/f}$ for initial prediction, and then upsamples the previous features and segmentation prediction to predict the next segmentation mask using the corresponding encoder features. This finally yields a four-level segmentation mask pyramid $\{S^1_{m/f}, S^2_{m/f}, S^3_{m/f}, S^4_{m/f}\}$ and a four-level segmentation feature pyramid $\{SF^1_{m/f}, SF^2_{m/f}, SF^3_{m/f}, SF^4_{m/f}\}$ that contains rich segmentation-related semantic information. Note that the segmentation decoder is used only in the semi-supervised setting where segmentation labels are partially available, while for unsupervised registration, this decoder is omitted.

### 2.2 Registration Decoder with Multi-level Multi-space Contextual Guidance

As the core component of M2CG, we design a registration decoder that integrates multi-level multi-space contextual guidance for one-pass progressive pyramid registration. It operates in a coarse-to-fine manner over four scale levels, leveraging four-space information, including the image features $IF^i_{m/f}$ extracted from the image encoder, the segmentation features $SF^i_{m/f}$ and masks $S^i_{m/f}$ derived from the segmentation decoder, and a four-level image pyramid $\{I^1_{m/f}, I^2_{m/f}, I^3_{m/f}, I^4_{m/f}\}$ obtained via wavelet transforms. The $I^1_{m/f}$ is the original input image $I_{m/f}$, while each of the $\{I^2_{m/f}, I^3_{m/f}, I^4_{m/f}\}$ denotes a 8-channel wavelet-decomposed image including 8 sub-bands of low-frequency approximation (*LLL*) and high-frequency coefficients (*LLH*, *LHL*, *HLL*, etc.).

At each scale level, the decode integrates contextual information from four spaces using four Intra-space Correlation Modules (IntraCMs). Each IntraCM captures spatial correspondence between the fixed and intermediately-warped images/masks/features via a 3D correlation layer followed by 3×3×3 convolutions. After the four IntraCMs is a Multi-space Integration Module (MultiIM) for fusing the correspondence information in four spaces and the upsampled features derived from its previous scale level. Each MultiIM consists of four 3×3×3 convolutions with dense-style residual connections to capture multi-scale deformations. Finally, a registration head is applied after MultiIM to predict the deformation field $\phi_i$ at this scale level, which is subsequently leveraged to warp $I^{i+1}_m$, $S^{i+1}_m$, $IF^{i+1}_m$, and $SF^{i+1}_m$ at the next scale level. In the unsupervised setting, the segmentation-related IntraCMs and operations are omitted.

The above process essentially incorporates three-level contextual guidance regarding the previous, current, and subsequent registration steps. The previous-level guidance refers to the use of upsampled features in MultiIM, providing contextual cues on how registration was performed at the previous scale level. The current-level guidance refers to employing four IntraCMs to explore in four spaces the current distance to the target alignment. The subsequent-level guidance refers to the use of wavelet-decomposed images in image-space IntraCM. The wavelet transform can decompose an image into 8 downsampled sub-bands without information loss; in other words, the 8 sub-bands can provide full information of the original-scale image. Intuitively, incorporating wavelet-decomposed images as image-space contextual cues enables the current registration step to access finer-scale details that would otherwise only be available at the subsequent registration step. This provides a form of "look-ahead" guidance, allowing for anticipating the finer-scale alignment requirements while operating at the current scale. Consequently, the registration at each scale level is informed not only by the current and previous contexts, but also by a preview of the next step's image details.

### 2.3 Unsupervised and Semi-supervised Learning

In the unsupervised setting, the learnable parameters $\theta$ are optimized via an unsupervised loss $\mathcal{L}_{uns}$ that does not require any labels. The $\mathcal{L}_{uns}$ is defined as $\mathcal{L}_{uns} = \mathcal{L}_{sim} + \sigma\mathcal{L}_{reg}$, where $\mathcal{L}_{sim}$ is an image similarity term that penalizes the differences between $I_{m\circ\phi}$ and $I_f$, while $\mathcal{L}_{reg}$ is a regularization term that encourages smooth and invertible transformations $\phi$. The $\sigma$ is a regularization parameter that is set as 1 by default. We used negative local normalized cross-correlation (NCC) as $\mathcal{L}_{sim}$, which is a commonly used similarity metric in deformable medical image registration. For $\mathcal{L}_{reg}$, we imposed a diffusion regularizer on $\phi$ to encourage its smoothness.

In the semi-supervised setting, the learnable parameters $\theta$ are optimized via a semi-supervised loss $\mathcal{L}_{semi}$ that requires segmentation labels $L_{m/f}$ available for partial training data. Following [19], the $\mathcal{L}_{semi}$ is defined as $\mathcal{L}_{semi} = \mathcal{L}_{uns} + \alpha\mathcal{L}_{seg} + \beta\mathcal{L}_{joint}$, where $\mathcal{L}_{seg}$ is a segmentation term that constrains the predicted segmentation masks $S_m$ and $S_f$, while the $\mathcal{L}_{joint}$ is a joint term that imposes mutual constraints between registration and segmentation. The $\alpha$ and $\beta$ are two balancing parameters that are set to be 1 by default. As detailed in [19], the sum of Dice [20] and Focal [21] losses was used in $\mathcal{L}_{seg}$ and $\mathcal{L}_{joint}$, encouraging the consistency between (i) $S_{m/f}$ and $L_{m/f}$, (ii) $L_f$ and $L_m \circ \phi$, (iii) $L_f$ and $S_m \circ \phi$, and (iv) $S_f$ and $L_m \circ \phi$. For partially labeled training data, when $L_{m/f}$ is unavailable, its related terms are excluded from calculation, allowing both labeled and unlabeled data to be used for training.

## 3 Experimental Setup

### 3.1 Datasets and Preprocessing

Our method was evaluated on inter-patient brain and intra-patient cardiac image registration, involving eight public datasets. We followed the dataset settings in the previous study [19]: For brain image registration, 414 segmentation-labeled brain MRI images from OASIS [22] were randomly split into 314, 20, and 80 for training, validation, and testing. Four unlabeled brain MRI datasets, ADNI [23], ABIDE [24], ADHD [25], and IXI [26], were used for training, establishing a large semi-supervised training set with 2,656 unlabeled and 314 labeled images. Another two segmentation-labeled brain MRI datasets, Mindboggle [27] and Buckner [28], were used for independent testing. For cardiac image registration, the ACDC [29] dataset was used following its official data-splitting. ACDC uses intra-patient ED-ES registration, containing substantial physiological deformation, which serves as a meaningful test for large-deformation registration. Refer to [19] for detailed dataset settings and preprocessing procedures.

### 3.2 Implementation Details

Experiments were implemented with PyTorch on an 40 GB NVIDIA A100 GPU. We used an ADAM optimizer with a learning rate of 0.0001 and a batch size of 1. For brain image registration, deep models were trained for 100,000 iterations in total. In the unsupervised setting, inter-patient training image pairs were randomly picked from the training set, while in the semi-supervised setting, the first 50,000 iterations used labeled image pairs as a warm start, followed by 50,000 iterations using semi-labeled image pairs that consists of a labeled image and an unlabeled image.

For cardiac image registration, deep models were trained for 50,000 iterations in total. In the unsupervised setting, the first 40,000 iterations used intra-patient image pairs randomly picked from all available cine-MRI frames, followed by another 10,000 iterations using image pairs of End-Diastole (ED) and End-Systole (ES) frames for optimizing ED-ES registration. In the semi-supervised setting, the first 40,000 iterations used semi-labeled image pairs consisting of a labeled ED/ES frame and an unlabeled non-ED/ES frame, followed by 10,000 iterations using ED/ES frames only.

### 3.3 Experimental Designs

Our M2CG was extensively compared to existing registration methods in both unsupervised and semi-supervised settings, including traditional optimization-based methods, state-of-the-art one-pass/recursive methods, and semi-supervised methods that jointly perform segmentation to enhance registration. Also, two ablation studies were conducted to (i) validate the individual effectiveness of each contextual cue in M2CG, and (ii) investigate whether adding recursion to our M2CG is still beneficial.

Standard evaluation metrics for deformable registration tasks were adopted [36]: the registration accuracy was evaluated using the Dice similarity coefficients (DSC) between segmentation labels, while the smoothness of spatial transformations was evaluated using the percentage (%) of Negative Jacobian Determinants (NJD).

## 4 Results and Discussion

Table 1 presents the quantitative comparison with existing registration methods. Traditional methods were conducted using cross-correlation as the similarity measure, while all deep registration methods were trained using the same losses (described in Section 2.3) for fair comparisons. Our M2CG achieved the highest DSCs across all evaluation datasets in both unsupervised and semi-supervised settings, demonstrating its generalizability across application scenarios. Recursive methods achieved overall higher DSCs than one-pass methods despite requiring overall longer runtime, which aligns with the prevailing view of recursion versus one-pass trade-off [17]. Nevertheless, our M2CG still outperformed recursive methods with only one-pass computational cost. Further, semi-supervised methods that can leverage segmentation semantics achieved overall higher DSCs than unsupervised methods, and the performance gaps between recursive and one-pass methods in the semi-supervised setting are smaller compared to the gaps in the unsupervised setting. This implies that incorporating segmentation semantics has already undermined the necessity of recursion. It also should be noted that our M2CG exhibited lower or similar NJDs across all datasets and both settings, suggesting that it did not sacrifice transformation smoothness for registration accuracy. In addition, Fig. 2 presents a qualitative comparison with existing methods, where the registration results produced by our M2CG are the most consistent with the fixed image.

**Table 1.** Quantitative comparison with existing registration methods.

| Method | OASIS | | Mindboggle | | Buckner | | ACDC | | GPU Runtime↓ | |
|---|---|---|---|---|---|---|---|---|---|---|
| | DSC↑ | NJD↓ | DSC↑ | NJD↓ | DSC↑ | NJD↓ | DSC↑ | NJD↓ | Brain | Cardiac |
| Before registration | 0.610 | / | 0.347 | / | 0.406 | / | 0.590 | / | / | / |
| *Traditional* | | | | | | | | | | |
| SyN [3] | 0.765 | 2.227 | 0.534 | 1.956 | 0.567 | 1.874 | 0.747 | 0.398 | / | / |
| NiftyReg [4] | 0.788 | 2.341 | 0.569 | 2.364 | 0.611 | 2.175 | 0.768 | 0.421 | / | / |
| *Uns, Recursive* | | | | | | | | | | |
| RCN [11] | $0.795^*$ | 2.985 | $0.592^*$ | 3.950 | $0.630^*$ | 4.020 | $0.785^*$ | 0.551 | 0.69s | 0.10s |
| LapIRN [12] | $0.802^*$ | 2.112 | $0.605^*$ | 2.164 | $0.632^*$ | 2.112 | $0.790^*$ | 0.454 | 0.46s | 0.06s |
| RDN [14] | $0.798^*$ | 1.996 | $0.599^*$ | 2.021 | $0.642^*$ | 1.932 | $0.798^*$ | 0.458 | 0.58s | 0.08s |
| RDP [15] | $0.805^*$ | 1.842 | $0.635^*$ | <u>1.798</u> | $0.656^*$ | <u>1.833</u> | $0.801^*$ | <u>0.378</u> | 0.79s | 0.12s |
| ReCorr [30] | <u>0.806</u>$^*$ | <u>1.793</u> | <u>0.637</u>$^*$ | 1.853 | <u>0.657</u>$^*$ | 1.873 | <u>0.803</u>$^*$ | 0.392 | 0.51s | 0.07s |
| *Uns, One-pass* | | | | | | | | | | |
| VoxelMorph [5] | $0.773^*$ | 1.997 | $0.552^*$ | 2.532 | $0.589^*$ | 2.220 | $0.754^*$ | 0.440 | **0.23s** | **0.02s** |
| TransMorph [6] | $0.792^*$ | 1.908 | $0.571^*$ | 2.400 | $0.608^*$ | 2.183 | $0.768^*$ | 0.492 | 0.35s | 0.05s |
| NICE-Net [7] | $0.796^*$ | 1.832 | $0.618^*$ | 2.043 | $0.643^*$ | 1.963 | $0.785^*$ | 0.443 | <u>0.32s</u> | <u>0.04s</u> |
| NICE-Trans [8] | $0.800^*$ | 1.897 | $0.625^*$ | 2.324 | $0.649^*$ | 2.277 | $0.799^*$ | 0.473 | 0.37s | 0.05s |
| SACB-Net [31] | $0.802^*$ | 2.531 | $0.632^*$ | 2.126 | $0.653^*$ | 1.992 | $0.800^*$ | 0.852 | 0.46s | 0.06s |
| M2CG (Ours, Uns) | **0.815** | **1.689** | **0.648** | **1.721** | **0.665** | **1.814** | **0.810** | **0.358** | <u>0.32s</u> | <u>0.04s</u> |
| *Semi, Recursive* | | | | | | | | | | |
| RDP‡ [15] | $0.845^†$ | 1.773 | $0.644^†$ | <u>1.744</u> | $0.667^†$ | <u>1.757</u> | $0.859^†$ | <u>0.356</u> | 0.79s | 0.12s |
| ReCorr‡ [30] | <u>0.847</u>$^†$ | 1.749 | $0.648^†$ | 1.779 | $0.667^†$ | 1.765 | <u>0.861</u>$^†$ | 0.364 | 0.51s | 0.07s |
| *Semi, One-pass* | | | | | | | | | | |
| GL-Net [32] | $0.828^†$ | <u>1.693</u> | $0.587^†$ | 1.984 | $0.629^†$ | 1.770 | $0.832^†$ | 0.385 | 0.62s | 0.08s |
| AutoFuse [19] | $0.842^†$ | 1.756 | $0.610^†$ | 2.115 | $0.652^†$ | 1.844 | $0.853^†$ | 0.462 | 0.59s | 0.08s |
| DAFF-Net [33] | $0.846^†$ | 1.880 | <u>0.654</u>$^†$ | 1.842 | <u>0.668</u>$^†$ | 1.811 | $0.859^†$ | 0.401 | <u>0.45s</u> | <u>0.06s</u> |
| M2CG (Ours, Semi) | **0.862** | **1.668** | **0.665** | **1.698** | **0.687** | **1.722** | **0.872** | **0.348** | **0.40s** | **0.05s** |

**Bold**/<u>Underlined</u>: The best/second-best in each setting. ↑/↓: higher/lower is better. */†: $P$<0.05, in comparison to unsupervised/semi-supervised M2CG. ‡: Trained under DeepAtlas [34].

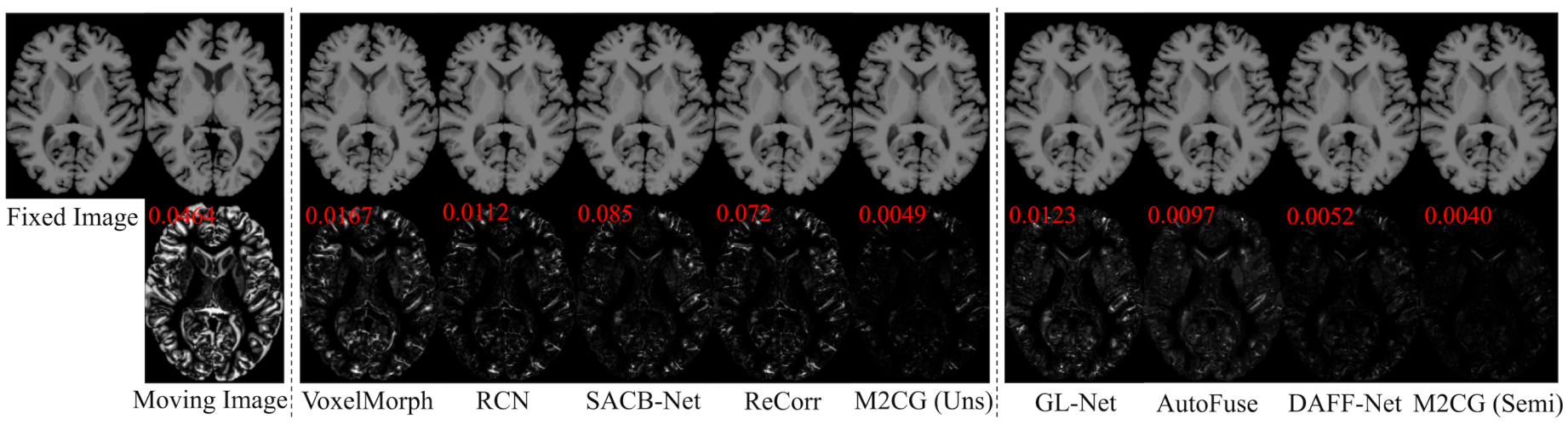

**Fig. 2.** Qualitative comparison with existing registration methods. Below each image is an error map that shows the intensity differences between the warped image and the fixed image, with the mean absolute error marked in red in the upper left corner.

Table 2 presents the results of ablation studies in the unsupervised/semi-supervised settings. First, we separately removed each contextual cue and observed consistent performance degradation in both settings, demonstrating the individual effectiveness of each cue in multi-level multi-space contextual guidance. Second, we separately added three recursion mechanisms into M2CG: (i) network recursion following RCN [11], (ii) decoder recursion following RDN [14], and (iii) pyramid-scale recursion following RDP [15]. We observed that adding recursion yielded only marginal improvements in the unsupervised setting and no improvement in the semi-supervised setting, which validates our hypothesis that the necessity of recursion decays with more informative contextual guidance provided. Due to GPU memory constraints, recursion was performed only twice, but this has already incurred almost twice the computational cost and GPU memory usage. Moreover, adding recursion adversely increased NJDs, suggesting that errors or transformation folds may accumulate with repeated registration.

**Table 2.** Ablation Studies in the unsupervised/semi-supervised settings.

| Method | OASIS | | Mindboggle | | Buckner | | ACDC | |
|---|---|---|---|---|---|---|---|---|
| | DSC↑ | NJD↓ | DSC↑ | NJD↓ | DSC↑ | NJD↓ | DSC↑ | NJD↓ |
| *Unsupervised* | | | | | | | | |
| M2CG (Ours, Uns) | 0.815 | **1.689** | 0.648 | **1.721** | 0.665 | **1.814** | **0.810** | **0.358** |
| w/o Contexts in image space | 0.803$^*$ | 1.774 | 0.630$^*$ | 1.784 | 0.651$^*$ | 1.937 | 0.796$^*$ | 0.387 |
| w/o Contexts in image feature space | 0.796$^*$ | 1.863 | 0.612$^*$ | 2.114 | 0.629$^*$ | 1.953 | 0.790$^*$ | 0.415 |
| w/o Contexts in previous level | 0.810$^*$ | 2.045 | 0.645 | 1.753 | 0.662 | 2.101 | 0.808 | 0.426 |
| w/o Contexts in subsequent level | 0.806$^*$ | 1.741 | 0.634$^*$ | 1.843 | 0.655$^*$ | 1.921 | 0.801$^*$ | 0.369 |
| Add Network Recursion | 0.815 | 2.533 | 0.647 | 3.124 | 0.665 | 3.234 | 0.809 | 0.742 |
| Add Decoder Recursion | 0.814 | 2.132 | 0.648 | 2.864 | 0.665 | 2.852 | 0.809 | 0.684 |
| Add Pyramid-scale Recursion | **0.816** | 2.376 | **0.650** | 2.732 | **0.666** | 2.637 | **0.810** | 0.627 |
| *Semi-supervised* | | | | | | | | |
| M2CG (Ours, Semi) | **0.862** | **1.668** | **0.665** | **1.698** | **0.687** | **1.722** | **0.872** | **0.348** |
| w/o Contexts in image space | 0.853$^\dagger$ | 1.785 | 0.658$^\dagger$ | 1.758 | 0.682$^\dagger$ | 1.747 | 0.865$^\dagger$ | 0.367 |
| w/o Contexts in image feature space | 0.845$^\dagger$ | 1.876 | 0.649$^\dagger$ | 2.860 | 0.669$^\dagger$ | 1.851 | 0.857$^\dagger$ | 0.373 |
| w/o Contexts in seg space | 0.851$^\dagger$ | 1.758 | 0.657$^\dagger$ | 1.947 | 0.675$^\dagger$ | 1.785 | 0.863$^\dagger$ | 0.385 |
| w/o Contexts in seg feature space | 0.846$^\dagger$ | 1.699 | 0.652$^\dagger$ | 1.788 | 0.670$^\dagger$ | 1.934 | 0.854$^\dagger$ | 0.398 |
| w/o Contexts in previous level | 0.860 | 1.963 | 0.663 | 1.723 | 0.685 | 1.950 | 0.870 | 0.402 |
| w/o Contexts in subsequent level | 0.855$^\dagger$ | 1.829 | 0.660$^\dagger$ | 1.959 | 0.684 | 1.849 | 0.866$^\dagger$ | 0.359 |
| Add Network Recursion | 0.861 | 2.436 | 0.664 | 3.325 | 0.685 | 3.542 | 0.869 | 0.823 |
| Add Decoder Recursion | 0.860 | 2.326 | 0.663 | 3.273 | 0.685 | 3.248 | 0.871 | 0.724 |
| Add Pyramid-scale Recursion | **0.862** | 2.583 | 0.664 | 2.976 | 0.686 | 3.117 | 0.871 | 0.773 |

**Bold**/Underlined: The best/second-best in each setting. ↑/↓: higher/lower is better. */†: $P$<0.05, in comparison to unsupervised/semi-supervised M2CG.

## 5 Conclusion

In this study, we have outlined **M2CG**, a one-pass pyramid registration network with **M**ulti-level **M**ulti-space **C**ontextual **G**uidance, to challenge the necessity of recursion in deformable medical image registration. By integrating contextual cues from three progressive levels and four complementary spaces, our M2CG achieves accurate registration within a single network iteration. Extensive experiments, involving eight public datasets across two inter-/intra-patient medical registration tasks, demonstrate state-of-the-art performance under both unsupervised and semi-supervised settings. Ablation analysis shows that adding recursion to M2CG provides limited accuracy gains but increases computational cost and transformation folding. Our findings suggest that the current reliance on recursion may stem from insufficient contextual guidance rather than inherent limitations of one-pass architectures. We emphasize that this conclusion is scoped to the evaluated uni-modal settings; validating M2CG on more anatomical sites, cross-modal registration, and clinically heterogeneous scenarios [37] remains an important direction for future work.

**Acknowledgments**. This work was supported by the Zhongguancun Academy (Grant No. XTS0071).

**Disclosure of Interests.** The authors have no competing interests to declare.